\documentclass[journal]{vgtc}                

\usepackage{mathptmx}
\usepackage{graphicx}
\usepackage{times}
\usepackage[mathscr]{euscript}
\usepackage{tabularx}
\usepackage[final]{review}
\usepackage{enumitem}
\usepackage{wasysym}
\usepackage{array}
\usepackage{tcolorbox}
\usepackage[switch]{lineno}
\usepackage{nameref}

\usepackage[bookmarks,backref=false,linkcolor=blue]{hyperref} 
\hypersetup{
  pdfauthor = {},
  pdftitle = {},
  pdfsubject = {},
  pdfkeywords = {},
  colorlinks=true,
  linkcolor= black,
  citecolor= black,
  pageanchor=true,
  urlcolor = black,
  plainpages = false,
  linktocpage
}

\onlineid{0}

\vgtccategory{Research}

\title{A Framework for Creative Visualization-Opportunities Workshops}

\author{
Ethan Kerzner,
Sarah Goodwin,
Jason Dykes,
Sara Jones,
Miriah Meyer
}
\authorfooter{
\item
 Ethan Kerzner and Miriah Meyer are with the University of Utah. \\ E-mail: kerzner@sci.utah.edu and miriah@cs.utah.edu.
\item
 Sarah Goodwin is with the Royal Melbourne Institute of Technology and Monash University. E-mail: sarah.goodwin@monash.edu.
\item
 Jason Dykes and Sara Jones are with City, University of London. \\ E-mail: [j.dykes,s.v.jones]@city.ac.uk.
}

\shortauthortitle{Kerzner \MakeLowercase{\textit{et al.}}: A Framework for Creativity Workshops in Applied Visualization Research}

\abstract{Applied visualization researchers often work closely with domain collaborators to explore new and useful applications of visualization. The early stages of collaborations are typically time consuming for all stakeholders as researchers piece together an understanding of domain challenges from disparate discussions and meetings. A number of recent projects, however, report on the use of creative visualization-opportunities (CVO) workshops to accelerate the early stages of applied work, eliciting a wealth of requirements in a few days of focused work. Yet, there is no established guidance for how to use such workshops effectively. In this paper, we present the results of a 2-year collaboration in which we analyzed the use of 17 workshops in 10 visualization contexts. Its primary contribution is a framework for CVO workshops that: 1) identifies a process model for using workshops; 2) describes a structure of what happens within effective workshops; 3) recommends \numberOfGuidelines actionable guidelines for future workshops; and 4) presents an example workshop and workshop methods. The creation of this framework exemplifies the use of critical reflection to learn about visualization in practice from diverse studies and experience.}


\CCScatlist{ 
 \CCScat{K.6.1}{Management of Computing and Information Systems}%
{Project and People Management}{Life Cycle};
 \CCScat{K.7.m}{The Computing Profession}{Miscellaneous}{Ethics}
}

\makeatletter
\newcommand{\customlabel}[2]{%
  \@bsphack
  \csname phantomsection\endcsname
  \def\@currentlabel{#1}{\label{#2}}%
  \@esphack
  \textbf{#1}
  }
\makeatother

\newcounter{procounter}
\renewcommand{\theprocounter}{P.\arabic{procounter}.}

\newcounter{reccounter}
\renewcommand{\thereccounter}{$\mathscr{R}$.\arabic{reccounter}}
\newcounter{descounter}
\renewcommand{\thedescounter}{$\mathscr{D}$.\arabic{descounter}}
\usepackage{xspace}
\newcommand{\Tactic}[2]{\fontfamily{qcr}\selectfont{(\textbf{#1})#2}\rmfamily\xspace}
\newcommand{\tactic}[1]{\fontfamily{qcr}\selectfont{#1}\rmfamily\xspace}
\def \topic {\tactic{topic}}

\def \agency {\tactic{agency}}
\def \collegiality {\tactic{collegiality}}
\def \trust {\tactic{trust}}
\def \interest {\tactic{interest}}
\def \challenge {\tactic{challenge}}

\def \tactics {\tactic{TACTICs}}

\def \numberOfPitfalls{25\xspace}
\def \numberOfGuidelines{25\xspace}
\def \numberOfAudits{30\xspace}
\def \numberOfExampleMethods{15\xspace}

\def \workshops{CVO workshops\xspace} 
\def \workshop{CVO workshop\xspace}

\def \stickyNotes{sticky notes\xspace}

\def \methods {methods\xspace}
\def \methodology {methodology\xspace}
\def \mindset {mindset\xspace}

\def \mindset {mindset\xspace}

\begin{document}
\firstsection{Introduction}
\maketitle

Two key challenges in the early stages of applied visualization research are to find pressing domain problems and to translate them into interesting visualization opportunities. Researchers often discover such problems through a lengthy process of interviews and observations with domain collaborators that can sometimes take months~\cite{Lam2012,McKenna2014,Sedlmair2012}. A number of recent projects, however, report on the use of workshops to characterize domain problems in just a few days of focused work~\cite{Dykes2010,Goodwin2013,Goodwin2016,Kerzner2017,Nobre2017,Walker2013}. More specifically, these workshops are {\bf creative visualization-opportunities workshops (\workshops)}, in which researchers and their collaborators explore opportunities for visualization in a domain~\cite{Goodwin2013}. When used effectively, such workshops reduce the time and effort needed for the early stages of applied visualization work, as noted by one participant: \emph{``The interpersonal leveling and intense revisiting of concepts made more progress in a day than we make in a year of lab meetings \ldots [the workshop] created consensus by exposing shared user needs''}~\cite{Kerzner2017}.

The \workshops reported in the literature were derived and adapted from software requirements workshops~\cite{Jones2007} and creative problem-solving workshops~\cite{CreativeEducationFoundation2015} to account for the specific needs of visualization design. These adaptations were necessary because existing workshop guidance does not appropriately emphasize three characteristics fundamental to applied visualization, which we term {\it visualization specifics}: the {\it visualization mindset} of researchers and collaborators characterized by a symbiotic collaboration~\cite{Sedlmair2012} and a deep and changing understanding of domain challenges and relevant visualizations~\cite{McCurdy2016a}; the connection to {\it visualization methodologies} that include process and design decision models~\cite{Munzner2009,Sedlmair2012}; and the use of {\it visualization methods} within workshops to focus on data analysis challenges and visualization opportunities~\cite{Goodwin2013}. 

The successful use of \workshops resulted from an ad hoc process in which researchers modified existing workshop guidance to meet the needs of their specific projects and reported the results in varying levels of detail. For example, Goodwin et al.~\cite{Goodwin2013} provide rich details, but with a focus on their experience using a series of workshops in a collaboration with energy analysts. In contrast, Kerzner et al.~\cite{Kerzner2017} summarize their workshop with neuroscientists in one sentence even though it profoundly influenced their research. Thus, there is currently no structured guidance about how to design, run, and analyze \workshops. Researchers who are interested in using such workshops must adapt and refine disparate workshop descriptions.

In this paper, we --- a group of visualization and creativity researchers who have been involved with every \workshop reported in the literature --- 
reflect on our collective experience and offer guidance about how and why to use \workshops in applied visualization. More specifically, this paper results from a 2-year international collaboration in which we applied a methodology of \emph{critically reflective practice}~\cite{Brookfield1998} to perform meta-analysis of our collective experience and research outputs from conducting 17 workshops in 10 visualization contexts~\cite{Dykes2010,Goodwin2016,Goodwin2013,Kerzner2017:utdb,Kerzner2017,Lisle2017,Nobre2017,Rogers2016,Rogers2017,Walker2013}, combined with a review of the workshop literature from the domains of design~\cite{Biskjaer2017,Dove2014,Kumar2012,Sanders2010}, software engineering~\cite{Horkoff2015,Jones2008,Jones2005,Jones2007,Maiden2010,Maiden2004,Maiden2005}, and creative problem-solving~\cite{DeBono1983,Gordon1961,Hamilton2016,Miller1989,Osborn1953}. 

This paper's primary contribution is a framework for \workshops. The framework consists of: 1) a process model that identifies actions before, during, and after workshops; 2) a structure that describes what happens in the beginning, in the middle, and at the end of effective workshops; 3) a set of \numberOfGuidelines actionable guidelines for future workshops; and 4) an example workshop and example methods for future workshops. To further enhance the actionability of the framework, in Supplemental Materials\footnote{\href{http://bit.ly/CVOWorkshops}{http://bit.ly/CVOWorkshops/}} we provide documents with expanded details of the example workshop, additional example methods, and \numberOfPitfalls pitfalls we have encountered when planning, running, and analyzing \workshops. 

We tentatively offer a secondary contribution: this work exemplifies critically reflective practice that enables us to draw upon multiple diverse studies to generate new knowledge about visualization in practice. Towards this secondary contribution we include, in Supplemental Materials, an {\it audit trail}~\cite{Carcary2009,Lincoln1985} of artifacts that shows how our thinking evolved over the 2-year collaboration. 

In this paper, we first summarize the motivation for creating this framework and describe related work in Sec.~\ref{sec:background} and \ref{sec:related}. Next, we describe our workshop experience and reflective analysis methods in Sec.~\ref{sec:experience} and ~\ref{sec:research}. Then, we introduce the framework in Sec.~\ref{sec:framework}~--~\ref{sec:after}. After that, we discuss implications and limitations of the work in Sec.~\ref{sec:discussion}. We conclude with future work in Sec.~\ref{sec:conclusion}.
\section{Motivation and Background}
\label{sec:background}

In our experience, \workshops provide tremendous value to applied visualization stakeholders --- researchers and the domain specialists with whom they collaborate. \workshops provide time for focused thinking about a collaboration, which allows stakeholders to share expertise and explore visualization opportunities. In feedback, one participant reported the workshop was {\it  ``a good way to stop thinking about technical issues and try to see the big picture''}~\cite{Goodwin2016}.

\workshops can also help researchers understand analysis pipelines, work productively within organizational constraints, and efficiently use limited meeting time. As another participant said: {\it ``The structured format helped us to keep on topic and to use the short time wisely. It also helped us rapidly focus on what were the most critical needs going forward. At first I was a little hesitant, but it was spot-on and wise to implement''}~\cite{Lisle2017}.

Furthermore, \workshops can build trust, rapport, and a feeling of co-ownership among project stakeholders. Researchers and collaborators can leave workshops feeling inspired and excited to continue a project, as reported by one participant: {\it ``I enjoyed seeing all of the information visualization ideas \ldots very stimulating for how these might be useful in my work''}~\cite{Goodwin2016}.

Based on these reasons, our view is that \workshops have saved us significant amounts of time pursuing problem characterizations and task analysis when compared to traditional visualization design approaches that involve one-on-one interviews and observations. What may have taken several months, we accomplished with several days of workshop preparation, execution, and analysis. In this paper we draw upon 10 years of experience using and refining workshops to propose a framework that enables others to use \workshops in the future.

\workshops are based on workshops used for software requirements and creative problem-solving~\cite{Goodwin2013}. Software requirements workshops elicit specifications for large-scale systems~\cite{Jones2007} that can be used in requirements engineering~\cite{Jones2005} and agile development~\cite{Hollis2013}. There are many documented uses of such workshops~\cite{Jones2008,Maiden2004,Maiden2007,Maiden2005}, but they do not appropriately emphasize the mindset of visualization researchers or a focus on data and analysis.

More generally, creative problem-solving workshops are used to identify and solve problems in a number of domains~\cite{Osborn1953} --- many frameworks exist for such workshops~\cite{CreativeEducationFoundation2015,DeBono1983,Gordon1961,Gray2010,Kumar2012}. Meta-analysis of these frameworks reveal common workshop characteristics that include: promoting trust and risk taking, exploring a broad space of ideas, providing time for focused work, emphasizing both problem finding and solving, and eliciting group creativity from the cross-pollination of ideas~\cite{Nickerson1999}. 


Existing workshop guidance, however, does not completely describe \workshops. The key distinguishing feature of \workshops is the explicit focus on visualization, which implies three {\bf visualization specifics} for effective workshops and workshop guidance:
\begin{itemize}[nolistsep,noitemsep]
    \item Workshops should promote a {\bf visualization mindset} --- the set of beliefs and attitudes held by project stakeholders, including an evolving understanding about domain challenges and visualization~\cite{McCurdy2016a,Sedlmair2012} --- that fosters and benefits an exploratory and visual approach to dealing with data while promoting trust and rapport among these stakeholders~\cite{Shneiderman2006};
    \item Workshops should contribute to {\bf visualization methodologies} --- the research practices of visualization, including process and decision models~\cite{McKenna2014,Munzner2009} --- by creating artifacts and knowledge useful in the visualization design process; and
    \item Workshops should use {\bf visualization methods} that explicitly focus on data visualization and analysis by exploring visualization opportunities with the appropriate {\it information location} and {\it task clarity}~\cite{Sedlmair2012}.
\end{itemize}
This paper is, in part, about adopting and adapting creative problem-solving workshops to account for these visualization specifics.

\section{Related Work}
\label{sec:related}

Workshops are commonly used in a number of fields, such as business~\cite{Gray2010,Hamilton2016,Stanfield2002} and education~\cite{Anderson2000,Brooks-Harris1999}. Guidance from these fields, however, does not emphasize the role of workshops in a design process, which is central to applied visualization. Therefore, we focus this section on workshops as visualization design methods.  

\workshops can be framed as a method for user-centered design~\cite{Norman1986}, participatory design~\cite{Muller1993}, or co-design~\cite{Sanders2008} because they involve users directly in the design process --- we draw on work from these fields that have characterized design methods. Sanders et al.~\cite{Sanders2010}, for example, characterize methods by their role in the design process. Biskjaer et al.~\cite{Biskjaer2017} analyze methods based on concrete, conceptual, and design space aspects. Vines et al.~\cite{Vines2013} propose ways of thinking about how users are involved in design. Dove~\cite{Dove2016} describes a framework for using data visualization in participatory workshops. A number of books also survey existing design methods~\cite{Buxton2010,Kumar2012} and  practices~\cite{Knapp2016,Laural2003,Sanders2013}. These resources are valuable for understanding design methods but do not account for visualization specifics such as methodologies that emphasize the critical role of data early in the design process~\cite{Lloyd2011}.

\workshops can also be framed within existing visualization design process and decision models~\cite{Marai2018,McKenna2014,Munzner2009,Sedlmair2012,Tory2004}. More specifically, \workshops focus on eliciting opportunities for visualization software from collaborators. They support the {\it understand} and {\it ideate} design activities~\cite{McKenna2014} or fulfill the {\it winnow}, {\it cast}, and {\it discover} stages of the design study methodology's nine-stage framework~\cite{Sedlmair2012}.

A number of additional methods can be used in the early stages of applied work. Sakai and Aert~\cite{Sakai2015}, for example, describe the use of card sorting for problem characterization. McKenna et al.~\cite{McKenna2015} summarize the use of qualitative coding, personas, and data sketches in collaboration with security analysts. Koh et al.~\cite{Koh2011} describe workshops that demonstrate a wide range of visualizations to domain collaborators, a method that we have adapted for use in \workshops as described in Sec.~\ref{sec:workshop-methods}. Roberts et al.~\cite{Roberts2016} describe a method for exploring and developing visualization ideas through structured sketching. This paper is about how to use these design methods, and others, within structured \workshops.

Visualization education workshops are also relevant to \workshops. Huron et al.~\cite{Huron2016} describe data physicalization workshops for constructive visualization with novices. He et al.~\cite{He2017} describe workshops for students to think about the relationships between domain problems and visualization designs. In contrast, we frame \workshops as a method for experienced researchers to pursue domain problem characterization. Nevertheless, we see opportunities for participatory methods, such as constructive visualization~\cite{Huron2014} and sketching~\cite{Walny2015}, to be integrated into \workshops.
\begin{table*}
    \small
    \centering
    \begin{tabular}{|lllllllll|}
        \hline
        \textbf{ID} & \textbf{Year} & \textbf{Domain} & \textbf{Summary} & \textbf{Workshops} & \textbf{Result} & \textbf{Prim.} & \textbf{Supp.} & \textbf{Ref.} \\
        \hline
        \customlabel{P1}{pro:edina} & 2009 & Cartography  & \emph{``“Reimagining the legend as an exploratory visualization interface”''} & 3 & Paper & JD & *  & \cite{Dykes2010} \\
        \customlabel{P2}{pro:eon} & 2012 & Smart Homes & Deliver insights into the role of smart homes and new business potential & 4 & Paper & SG & JD,SJ,* & \cite{Goodwin2013}\\
        \customlabel{P3}{pro:htva} & 2012 & Human terrain & \emph{``develop [visualization] techniques that are meaningful in HTA''} & 3 & Paper & JD & * & \cite{Walker2013}\\
        \customlabel{P4}{pro:graffinity} & 2015 & Neuroscience & Explore problem-driven multivariate graph visualization & 1 &  Paper & EK & MM, * & \cite{Kerzner2017} \\ 
        \customlabel{P5}{pro:cp} & 2015 & Constraint prog. & Design performance profiling methods for constraint programmers & 1 & Paper & SG & * & \cite{Goodwin2016}\\
        \customlabel{P6}{pro:lineage} & 2017 & Psychiatry & Support visual analysis of determining or associated factors of suicide & 1 & Paper & * & EK,* & \cite{Nobre2017}\\ 
        \customlabel{P7}{pro:updb} & 2017 & Genealogy  & Discover opportunities to support visual genealogy analysis & 1 &  ---  & * & EK,MM,* & \cite{Kerzner2017:utdb}\\
        \customlabel{P8}{pro:arbor} & 2017 & Biology & Support phylogenetic analysis with visualization software & 1 &  In-progress & * & EK,MM,*  & \cite{Lisle2017}\\
        \hline
    \end{tabular}
    \caption{Summary of the projects in which we have used \workshops: six resulted in publications [\ref{pro:edina} -- \ref{pro:lineage}], one did not result in active collaboration  [\ref{pro:updb}], and one is in progress [\ref{pro:arbor}]. We characterize our involvement in these projects as either the primary researcher or as supporting researchers. The * represents colleagues who were involved in each project but who are not coauthors of this paper.}
    \label{tab:projects}
\end{table*}
\begin{table}
    \small
    \centering
    \begin{tabular}{|lp{4cm}lll|}
        \hline
        \textbf{ID} & \textbf{Theme} & \textbf{Facil.} & \textbf{Partic.} & \textbf{Hrs} \\ \hline
        {\bf \ref{pro:edina}} & Explore possibilities for enhancing legends with visualizations & 1v & 3v / 5c & 6 \\
        \hline
        {\bf \ref{pro:eon}} & Identify future opportunities for utilising smart home data/technologies & 2v / 1p & 0v / 5c & 6 \\
        \hline
        {\bf \ref{pro:htva}} & Identify novel visual approaches most suitable for HTA & 1v / 1p & 7v / 6c  & 9  \\
        \hline
        {\bf \ref{pro:graffinity}} & Explore shared user needs for visualization in retinal connectomics & 4v & 0v / 9c & 7 \\
        \hline
        {\bf \ref{pro:cp}} & Identify analysis and vis. opportunities for improved profiling of cons. prog. & 2v / 1c & 0v / 10c & 7 \\
        \hline
        {\bf \ref{pro:lineage}} & Understand the main tasks of psychiatric researchers & 2v & 1v / 6c & 3  \\
        \hline
        {\bf \ref{pro:updb}} & Explore opportunities for a design study with genealogists & 1v & 3v / 7c  & 3 \\
        \hline
        {\bf \ref{pro:arbor}} & Explore opportunities for funded collaboration between vis. and bio. & 1v / 1c & 2v / 12c & 7x2 \\
        \hline
    \end{tabular}
    \caption{Summary of the \workshop used in each project. We describe workshops by their theme, a concise statement the topics explored. We characterize workshop stakeholders as facilitators or participants categorized by their affiliation as (v)isualization researchers, (c)ollaborators, or (p)rofessional facilitators. Our workshops included 5 -- 14 participants and ranged in length from half a day to 2 days.}
    \label{tab:workshops}
\end{table}

\section{Workshop Experience and Terminology}
\label{sec:experience}

To develop the \workshop framework proposed in this paper, we gathered researchers who used workshops on 3 continents over the past 10 years. Our collective experience includes 17 workshops in 10 contexts: 15 workshops in 8 applied collaborations, summarized in Table~\ref{tab:projects} and Table~\ref{tab:workshops}; and 2 participatory workshops at IEEE VIS that focused on creating visualizations for domain specialists~\cite{Rogers2016,Rogers2017}.

The ways in which we use workshops have evolved over 10 years. In three of our projects, we used a series of workshops to explore opportunities, develop and iterate on prototypes, and evaluate the resulting visualizations in collaborations with cartographers~\cite{Dykes2010}, energy analysts~\cite{Goodwin2013}, and defense analysts~\cite{Walker2013}. In three additional projects, we used a single workshop to jump-start applied collaborations with neuroscientists~\cite{Kerzner2017}, constraint programmers~\cite{Goodwin2016}, and psychiatrists~\cite{Nobre2017}.  Recently, we used two workshops to explore opportunities for funded collaboration with genealogists~\cite{Kerzner2017:utdb} and biologists~\cite{Lisle2017}.

In our meta-analysis, we focused on the workshops used in the early stages of applied work or as the first in a series of workshops. To describe these workshops, we developed the term \workshops because they aim to deliberately and explicitly foster creativity while exploring opportunities for applied visualization collaborations.

Focused on \workshops, our experience includes the eight workshops in Table~\ref{tab:workshops}. Since we analyzed more data than appeared in any resulting publications, including artifacts and experiential knowledge, we refer to workshops and their projects by identifiers, e.g., [\ref{pro:edina}] refers to our collaboration with cartographers. In projects where we used more than one workshop [\ref{pro:edina} -- \ref{pro:htva}], the identifier corresponds to the {\it first} workshop in the series, unless otherwise specified. 

To describe our experience, we developed terminology for the role of researchers involved in each project. The {\bf primary researcher} is responsible for deciding to use a \workshop, executing it, and integrating its results into a collaboration. Alternatively, {\bf supporting researchers} provide guidance and support to the primary researcher. We have been involved with projects as both primary and supporting researchers (see Table~\ref{tab:projects}).  

We also adopt terminology to describe \workshops. Workshops are composed of {\bf methods} --- specific, repeatable and modular activities~\cite{Crotty1998}. The methods are designed around a {\bf theme} that identifies the workshop's central topic or purpose~\cite{Brooks-Harris1999}. The {\bf facilitators} plan and guide the workshop, and the {\bf participants} carry out the workshop methods. Typically the facilitators are visualization researchers and participants are domain collaborators, but, visualization researchers can participate [\ref{pro:edina},~\ref{pro:htva}], and collaborators can facilitate [\ref{pro:cp},~\ref{pro:arbor}]. We adopted and refined this vocabulary during our reflective analysis.
\section{Research Process}
\label{sec:research}

The contributions in this paper arise from {\it reflection} --- the analysis of experiences to generate insights~\cite{Boud1985,Schon1988}. More specifically, we applied a methodology of {\it critically reflective practice}~\cite{Brookfield1998}, summarized by Thompson and Thompson~\cite{Thompson2008} as {\it ``synthesizing experience, reflection, self-awareness and critical thinking to modify or change approaches to practice.''}

We analyzed our collective experience and our \workshop data, which consisted of documentation, artifacts, participant feedback, and research outputs. The analysis methods that we used can be described through three metaphorical lenses of critically reflective practice:
\begin{itemize}[nolistsep,noitemsep]
\item The lens of our collective experience --- we explored and articulated our experiential knowledge through interviews, discussions, card sorting, affinity diagramming, observation listing, and observations-to-insights~\cite{Kumar2012}. We codified our experience, individually and collectively, in both written and diagram form. We iteratively and critically examined our ideas in light of workshop documentation and artifacts.
\item The lens of existing theory --- we grounded our analysis and resulting framework in the literature of creativity and workshops~\cite{CreativeEducationFoundation2015,Biskjaer2017,DeBono1983,Gordon1961,Hamilton2016,Miller1989,Nickerson1999,Osborn1953,Sawyer2003,Sawyer2006,Shneiderman2005} as well as visualization design theory~\cite{McKenna2014,Munzner2009,Sedlmair2010,Tory2004}. 
\item The lens of our learners (i.e., readers) --- in addition to intertwining our analysis with additional workshops, we shared drafts of the framework with visualization researchers, and we used their feedback to make the framework more actionable and consistent.
\end{itemize}

Our reflective analysis, conducted over two years, was messy and iterative. It included periods of focused analysis and writing, followed by reflection on what we had written, which spurred additional analysis and rewriting. Throughout this time, we generated diverse artifacts, including models for thinking about how to use workshops, written reflections on which methods were valuable to workshop success, and collaborative writing about the value of workshops. This paper's Supplemental Material contains a timeline of significant events in our reflective analysis and \numberOfAudits supporting documents that show how our ideas evolved into the following framework.
\section{Fundamentals of the Framework}
\label{sec:framework}

The framework proposed in this paper describes how and why to use \workshops. We use the term {\it framework} because what we have created provides an interpretive understanding and approach to practice instead of causal or predictive knowledge~\cite{Jabareen2008}. 
The framework is a thinking tool to navigate the process of planning, running, and analyzing a workshop, but we note that it cannot resolve every question about workshops because the answers will vary with local experience, preference, and context. In this section, we describe a set of factors that contribute to workshop effectiveness, as well as introduce the workshop process model and structure. 
We intend for the framework to be complemented by existing workshop resources from outside of visualization~\cite{CreativeEducationFoundation2015,Brooks-Harris1999,Gray2010,Hamilton2016}.

\subsection{Tactics for Effective Workshops}

Reflecting on our experience and reviewing the relevant literature~\cite{Nickerson1999,Osborn1953,Sawyer2003,Sawyer2006,Shneiderman2005} enabled us to identify several key factors that contribute to the effectiveness of workshops: focusing on the \topic of visualization, data and analysis, while fostering, maintaining, and potentially varying the levels of \agency, \collegiality, \trust, \interest, and \challenge associated with each. We term these factors {\bf \tactics for effective workshops}:  
\begin{itemize}[noitemsep,nolistsep]
\item \Tactic{T}{opic} --- the space of ideas relevant to data, visualization, and domain challenges in the context of the workshop theme.
\item \Tactic{A}{gency} --- the sense of stakeholder ownership in the workshop, the workshop outcomes, and the research collaboration.
\item \Tactic{C}{ollegiality} --- the degree to which communication and collaboration occur among stakeholders.
\item \Tactic{T}{rust} -- the confidence that stakeholders have in each other, the workshop, the design process, and the researchers' expertise.
\item \Tactic{I}{nterest} --- the amount of attention, energy, and engagement to workshop methods by the stakeholders.
\item \Tactic{C}{hallenge} --- the stakeholders' barrier of entry to, and likelihood of success in, workshop methods.
\end{itemize}

The \tactics are not independent, consistent, or measurable. The extent to which they are fostered depends upon the context in which they are used, including various characteristics of the workshop --- often unknown in advance, although perhaps detectable by facilitators. Yet, selecting methods to maintain appropriate levels of \agency, \interest, and \trust ~--- while varying levels of \challenge and approaching the \topic from different perspectives --- likely helps workshops to have a positive influence on the \mindset of stakeholders and to generate ideas that move forward the \methodology of the project. Hence, we refer to the \tactics throughout this framework.

\subsection{Process Model and Structure}
\label{sec:process-and-structure}

\begin{figure}
\includegraphics[width=\columnwidth]{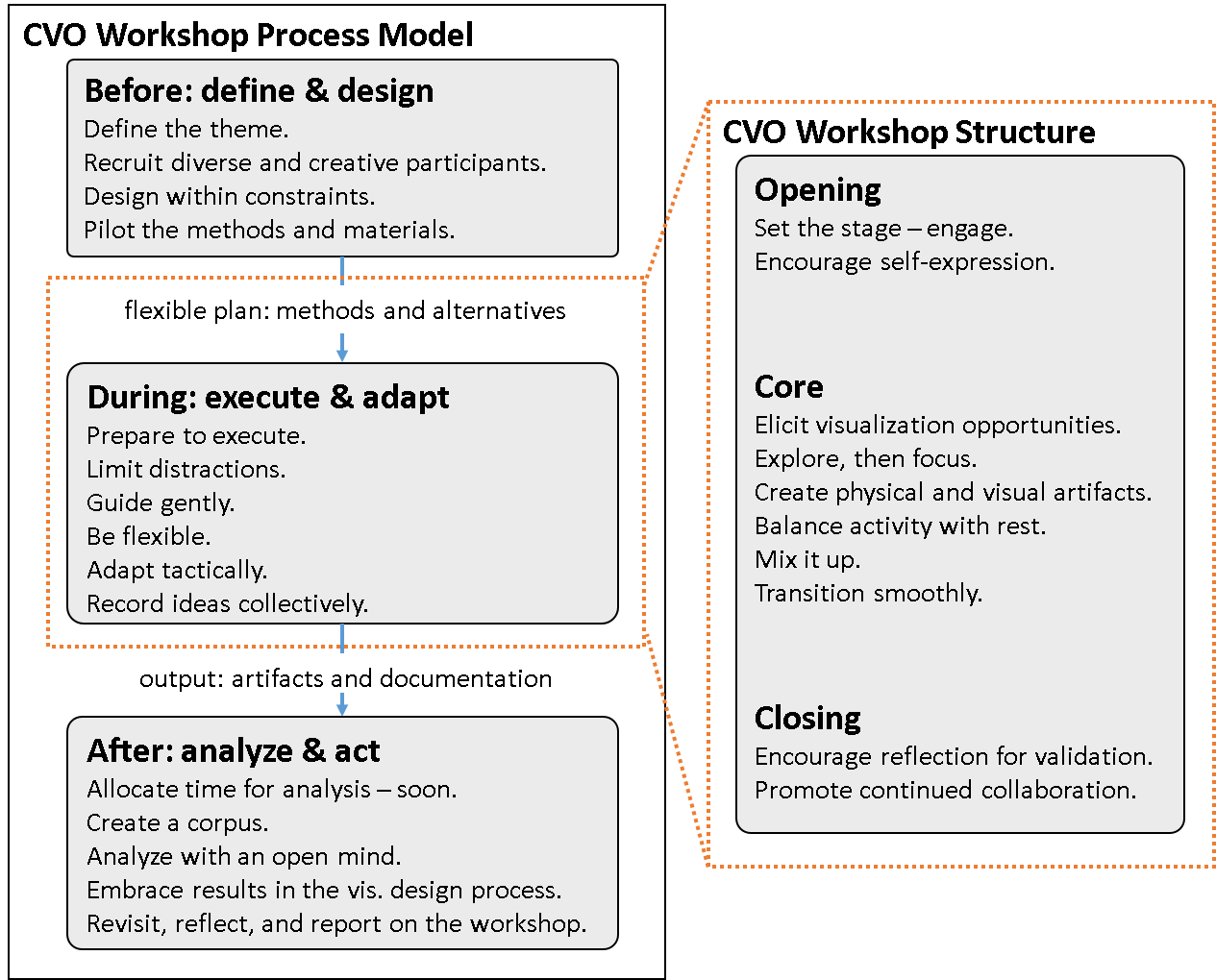}
\caption{The framework's two models are 1) a process model (left) that describes the common actions before, during, and after workshops; and 2) a structure that describes principles for methods used in the beginning, in the middle, and at the end of workshops. In these models, we propose \numberOfGuidelines guidelines for future workshops, summarized here.}
\label{fig:framework-overview}
\end{figure}

The framework proposes two models for describing how to use \workshops: a process model and a workshop structure. The models were adapted from the extensive literature that describes how to use workshops outside of visualization~\cite{CreativeEducationFoundation2015,Brooks-Harris1999,DeBono1983,Dove2016,Gray2010,Hamilton2016,Osborn1953}. 

The process model shown in Fig.~\ref{fig:framework-overview} (left) consists of three stages that describe the actions of using \workshops:
\begin{enumerate}[noitemsep,nolistsep]
    \item {\bf Before: define \& design.} Define the workshop theme and design workshop methods, creating a flexible workshop plan.
    \item {\bf During: execute \& adapt.} Perform the workshop plan, adapting it to participants' reactions in light of the \tactics, generating workshop output as a set of artifacts and documentation.
    \item {\bf After: analyze \& act.} Make sense of the workshop output and use it in the downstream design process.
\end{enumerate}

Nested within the process is the \workshop structure --- Fig.~\ref{fig:framework-overview} (right) --- that identifies key aspects of the methods used in the beginning, middle, and end of workshops:
\begin{enumerate}[nolistsep,noitemsep]
    \item {\bf Opening.} Establish shared context and \interest while promoting \trust, \agency, and \collegiality.
    \item {\bf Core.} Promote creative thinking about the \topic, potentially varying \challenge to maintain \interest.
    \item {\bf Closing.} Provide time for reflection on the \topic and promote continued \collegiality in the collaboration.
\end{enumerate}

The process model and structure are closely connected as shown by the orange box in Fig.~\ref{fig:framework-overview}. As part of the workshop process, we design and execute a workshop plan. This plan follows the workshop structure because it organizes methods into the opening, core, and closing. In other words, the process is about how we use a workshop; the structure is about how methods are organized within a workshop. 

We use the process model and structure to organize the following four sections of this paper. In these sections, we use paragraph-level headings to summarize \numberOfGuidelines actionable workshop guidelines. Additionally, in Supplemental Materials we include a complementary set of \numberOfPitfalls pitfalls that are positioned against these guidelines and the \tactics to further enhance the actionability of the framework.

\section{Before the Workshop: Define \& Design}
\label{sec:before}

Creating an effective \workshop is a design problem: there is no single correct workshop, the ideal workshop depends on its intended outcomes, and the space of possible workshops is practically infinite. Accordingly, workshop design is an iterative process of defining a goal, testing solutions, evaluating their effectiveness, and improving ideas. The framework we have developed here is part of this process. In this section, we introduce four guidelines --- summarized in paragraph-level headings --- for workshop design.

\paragraph{Define the theme.} Just as design starts with defining a problem, creating a \workshop starts with defining its purpose, typically by articulating a concise theme.
An effective theme piques \interest in the workshop through a clear indication of the \topic. It encourages a \mindset of mutual learning among stakeholders. It also focuses on opportunities that exhibit the appropriate {\it task clarity} and {\it information location} of the design study methodology~\cite{Sedlmair2012}. Examples from our work emphasize visualization opportunities (e.g., {\it ``enhancing legends with visualizations''} [\ref{pro:edina}]), domain challenges (e.g., {\it ``identify analysis and visualization opportunities for improved profiling of constraint programs''} [\ref{pro:cp}]), or broader areas of mutual interest (e.g., {\it``explore opportunities for a funded collaboration with phylogenetic analysts"}~[\ref{pro:arbor}]).

Although we can improve the theme as our understanding of the domain evolves, posing a theme early can ground the design process and identify promising participants.

\paragraph{Recruit diverse and creative participants.} \label{par:participants} We recruit participants who have relevant knowledge and diverse perspectives about the \topic. We also consider their openness to \challenge and potential \collegiality.

Examples of effective participants include a mix of frontline analysts, management, and support staff [\ref{pro:graffinity}]; practitioners, teachers, and students [\ref{pro:cp}]; or junior and senior analysts [\ref{pro:lineage}]. We recommend that participants attend the workshop in person because remote participation proved distracting in one workshop~[\ref{pro:arbor}]. Recruiting {\it fellow-tool builders}~\cite{Sedlmair2012} as participants should be approached with caution because their perspectives may distract from the \topic~--- this happened in our workshop that did not result in active collaboration [\ref{pro:updb}].

\paragraph{Design within constraints.} Identifying constraints can help winnow the possibilities for the workshop. Based on our experience, the following questions are particularly useful for workshop design:

\begin{itemize}[nolistsep,noitemsep]

\item Who will use the workshop results? Identifying the primary researcher early in the process is important because he or she will be responsible for the workshop and ultimately use its results. In a workshop where we did not clearly identify the primary researcher, the results went unused [\ref{pro:updb}].

\item How many participants will be in the workshop? We typically recruit 5 to 15 participants --- a majority domain collaborators, but sometimes designers and researchers [\ref{pro:edina},~\ref{pro:htva},~\ref{pro:lineage}~--~\ref{pro:arbor}].

\item Who will help to facilitate the workshop? We have facilitated our workshops as the primary researcher, with the assistance of supporting researchers or professional workshop facilitators. Domain collaborators can also be effective facilitators, especially if the domain vocabulary is complex and time is limited [\ref{pro:cp},~\ref{pro:arbor}].

\item How long will the workshop be? Although we have run workshops that range from half a day [\ref{pro:lineage},~\ref{pro:updb}] to two days [\ref{pro:arbor}], these extremes either feel rushed or require significant commitment from collaborators. We recommend that an effective workshop lasts about one working day.

\item Where will the workshop be run? Three factors are particularly important for determining the workshop venue: a mutually convenient location, a high quality projector for visualization examples, and ample space to complete the methods. We have had success with workshops at offsite locations [\ref{pro:eon},~\ref{pro:htva}], our workplaces, and our collaborators' workplaces [\ref{pro:graffinity}~--~\ref{pro:lineage}].

\item What are additional workshop constraints? Examples include the inability of collaborators to share sensitive data [\ref{pro:htva},~\ref{pro:lineage}] and the available funding.

\end{itemize} 

\paragraph{Pilot the methods and materials.} Piloting methods can ensure that the workshop will generate ideas relevant to the \topic while maintaining appropriate levels of \interest and \challenge. We have piloted methods to evaluate how understandable they are [\ref{pro:eon},~\ref{pro:graffinity}], to test whether they create results that can be used to advance visualization design methodologies [\ref{pro:lineage},~\ref{pro:arbor}], to find mistakes in method prompts [\ref{pro:eon},~\ref{pro:graffinity},~\ref{pro:lineage},~\ref{pro:arbor}], and to ensure that the materials are effective --- e.g., \stickyNotes are the correct size and visualizations are readable on the projector.

It is also useful to pilot workshops with proxy participants, such as researchers~[\ref{pro:graffinity}] or collaborators~[\ref{pro:arbor}]. Feedback from collaborators during pilots has helped us revise the theme, identify promising participants, and refine the workshop methods.

\section{Workshop Structure and Methods}
\label{sec:design}

This section describes guidelines for the methods used in the three phases of the \workshop structure (described in Sec.~\ref{sec:process-and-structure}) --- the opening, core, and closing. It concludes with a summary of an example workshop and resources for additional workshop methods.

\subsection{Workshop Opening}

The workshop opening communicates the goals and guidelines for participants, but it can be more than that. It can foster \agency by encouraging self-expression and idea generation. It can encourage \collegiality and \trust by promoting open communication, acknowledging expertise, and establishing a safe co-owned environment. It can also garner \interest by showing that the workshop will be useful and enjoyable. Two guidelines contribute to an effective opening.

\paragraph{Set the stage --- engage.} \workshops typically open with a short introduction that reiterates the theme and establishes shared context for participants and facilitators. We have introduced workshops as \emph{``guided activities that are meant to help us understand: what would you like to do with visualization?''}~[\ref{pro:graffinity}]. We have also used graphics that summarize the goals of our project, potentially priming participants to engage with the \topic of visualization [\ref{pro:htva}].

The opening can establish principles for creativity~\cite{CreativeEducationFoundation2015,Osborn1953}, potentially fostering \trust and \collegiality. We used the following principles in one of our workshops [\ref{pro:eon}]: 1) all ideas are valid, express and record them; 2) let everyone have their say; 3) be supportive of others; 4) instead of criticizing, create additional ideas; 5) think `possibility' -- not implementation; 6) speak in headlines and follow with detail; and 7) switch off all electronic devices.

Introduction presentations should be kept short to maintain \interest. Passive methods, such as lectures and presentations, can discourage participation at the outset. For example, we started one workshop [\ref{pro:arbor}] with a presentation on the current state of analysis tools. This presentation encouraged participants to passively listen rather than actively explore, establishing a passive mindset that we had to overcome in subsequent methods. An effective opening engages participants.

\paragraph{Encourage self-expression.} We use methods that encourage self-expression to support interpersonal leveling and to act on the creativity principles --- {\it all ideas are valid} and {\it be supportive of others}. Such interpersonal methods help to establish an atmosphere of \trust and \collegiality among participants and facilitators. They can also provide participants with a sense of \agency~\cite{Brooks-Harris1999}.

We have used interpersonal methods that ask participants to sketch ideas while suspending judgment~\cite{Rogers2017} or to introduce themselves through analogies as a potential primer for creativity (see analogy introduction in Sec.~\ref{sec:workshop-methods}). Overall, we use interpersonal methods in the opening to engage participants and facilitators, preparing them for the workshop core.

\subsection{Workshop Core}

In the workshop core, we harness the active and engaged mindset of participants by encouraging them to explore a wide ideaspace before selecting the more promising ideas. The methods in the core potentially generate hundreds of sticky notes, sketches, and other artifacts. Analysis of our experience and relevant literature leads us to suggest five guidelines for an effective core.

\paragraph{Elicit visualization opportunities.} We select workshop methods relevant to the \topic, asking participants about their current analysis challenges, limitations of existing tools, characteristics of their data, or the ways in which they would like to use visualization. This can be achieved by adding a visualization twist to existing design and workshop methods. 

In one workshop [\ref{pro:htva}], for example, we used a method that {\it ``developed user stories, considered relevant datasets, discussed alternative scenarios and sketched solutions"} with our domain collaborators. In retrospect, this method connected the \topic into a more general workshop method, user stories~\cite{Kumar2012}.

\paragraph{Explore, then focus.} We organize the core to first generate ideas using divergent methods that expand the ideaspace. Then, we evaluate ideas using convergent methods that winnow the ideaspace~\cite{Osborn1953}. Using divergent methods early in the core allows us to consider many possibilities while also promoting \agency and maintaining \interest. Then, convergent methods can narrow the ideaspace to the more promising ideas. 

Classifying methods as either divergent or convergent risks oversimplification as individual methods often include both divergent and convergent aspects. Consider our use of brainstorming~\cite{Osborn1953} during one workshop~[\ref{pro:edina}]: we asked participants to record \emph{``problems and successes associated with the current clients on \stickyNotes''} (divergent) and then to share the more interesting ideas (convergent). We classify this method as divergent because it creates ideas, despite the convergent discussion. In contrast, a convergent method may only involve grouping \stickyNotes from previous methods. Overall, in line with existing workshop guidance~\cite{CreativeEducationFoundation2015,DeBono1983,Hamilton2016,Osborn1953}, we judge methods by their intended impact on the ideaspace and organize the core with phases of divergent and convergent methods.

\paragraph{Create physical and visual artifacts.} We select methods by how they encourage participants to write, draw, or otherwise externalize their ideas. Externalizing ideas creates artifacts for us to analyze after the workshop. It aids creative thinking because expressing an idea forces the creator to elaborate it~\cite{Sawyer2006}, and promotes idea sharing that encourages \collegiality.

We consider the artifact materials to be important. Sticky notes are particularly useful because they enable participants to group or rank ideas and potentially to discover emergent concepts in the ideaspace~\cite{Dove2016}. We have used \stickyNotes in almost all of our workshops, often using their color to encode information about which method generated an idea, and their positions to relate, differentiate, or rank ideas. This can help establish consensus. It can also aid post-workshop analysis by recording how ideas evolved and were valued throughout the workshop. Additional materials effective for externalizing ideas include handouts with structured prompts, butcher paper, and poster boards. Using whiteboards is tempting, but ideas are easily lost if the boards are erased.

We also consider the form of ideas to be important. Effective methods create artifacts relevant to the theme and \topic of visualization. This can be achieved through the use of visual language (see wishful thinking in Sec.~\ref{sec:workshop-methods}) and by encouraging participants to sketch or draw, such as in storyboarding [\ref{pro:eon},~\ref{pro:graffinity},~\ref{pro:cp}]. We see many opportunities to create visual artifacts using existing methods, such as sketching with data~\cite{Walny2015}, constructive visualizations~\cite{Huron2014}, or parallel prototyping~\cite{Roberts2016} approaches.

\paragraph{Balance activity with rest.} Because continuously generating or discussing ideas can be tiring for participants, we structure workshop methods to provide a balance between activity and rest. Specifically, we incorporate passive methods that provide time for incubation, the conscious and unconscious combination of ideas~\cite{Sawyer2006}. 

Passive methods can include short breaks with food and coffee, informal discussions over meals, or methods where participants listen to presentations. When using methods that present ideas, asking participants to record their thoughts and reactions can promote \interest and maintain a feeling of \agency. We have typically used passive methods in full-day workshops [\ref{pro:eon},~\ref{pro:graffinity},~\ref{pro:cp},~\ref{pro:arbor}], but we rely on breaks between methods for shorter workshops [\ref{pro:lineage}].

\paragraph{Mix it up.} We consider the relationships among methods to be important as we strive to balance exploration with focus and activity with rest, while also using many materials for externalizing ideas. Considering methods that vary these factors can provide different levels of \challenge because, for example, methods that require drawing ideas may be more difficult than discussing ideas. Using a variety of methods may also maintain \interest because participants may become bored if too much time is spent on a specific idea.

\paragraph{Transition smoothly.} We avoid potentially jarring transitions between methods to preserve participant \interest. Convergent discussions can be used to conclude individual methods by highlighting the interesting, exciting, or influential ideas. These discussions can promote \collegiality by encouraging communication of ideas, \agency by validating participants' contributions, and \interest in the ideas generated. Convergent discussions also highlight potentially important ideas for researchers to focus on after the workshop.

Convergent methods can also conclude the workshop core by grouping or ranking key ideas. We have used storyboarding to encourage the synthesis of ideas into a single narrative [\ref{pro:eon},~\ref{pro:graffinity},~\ref{pro:cp}]. We have also asked participants to rank ideas, providing cues for analyzing the workshop results [\ref{pro:htva}]. Convergent methods provide a sense of validation, potentially helping to build \trust among researchers and collaborators as we transition to the closing. 

\subsection{Workshop Closing}

The workshop closing sets the tone for continued collaboration. It is an opportunity to promote \collegiality by reflecting on the shared creative experience. It allows for analysis that can potentially identify the more interesting visualization opportunities. The following two guidelines apply to effective closings. 

\paragraph{Encourage reflection for validation.} We use discussions at the end of workshops to encourage reflection, potentially providing validation to participants and generating information valuable for workshop analysis. We encourage participants to reflect on how their ideas have evolved by asking, \emph{``What do you know now that you did not know this morning?''} [\ref{pro:cp}] or \emph{``What will you do differently tomorrow, given what you have learned today?''} [\ref{pro:eon}]. Responses to these questions can provide validation for the time committed to the workshop. One participant, for example, reported, \emph{``I was surprised by how much overlap there was with the challenges I face in my own work and those faced by others''} [\ref{pro:cp}]. 


\paragraph{Promote continued collaboration.} We conclude the workshop by identifying the next steps of action --- continuing the \methodology of the collaboration. We can explain how the ideas will be used to move the collaboration forward, often with design methods as we describe in Sec.~\ref{sec:after}. 

We can also ask participants for feedback about the workshop to learn more about their perceptions of visualization and to evaluate the effectiveness of workshop methods --- encouraging the visualization mindset. E-mailing online surveys immediately after a workshop is effective for gathering feedback [\ref{pro:graffinity},~\ref{pro:arbor}].

\begin{figure}
    \centering
    \includegraphics[width=\columnwidth]{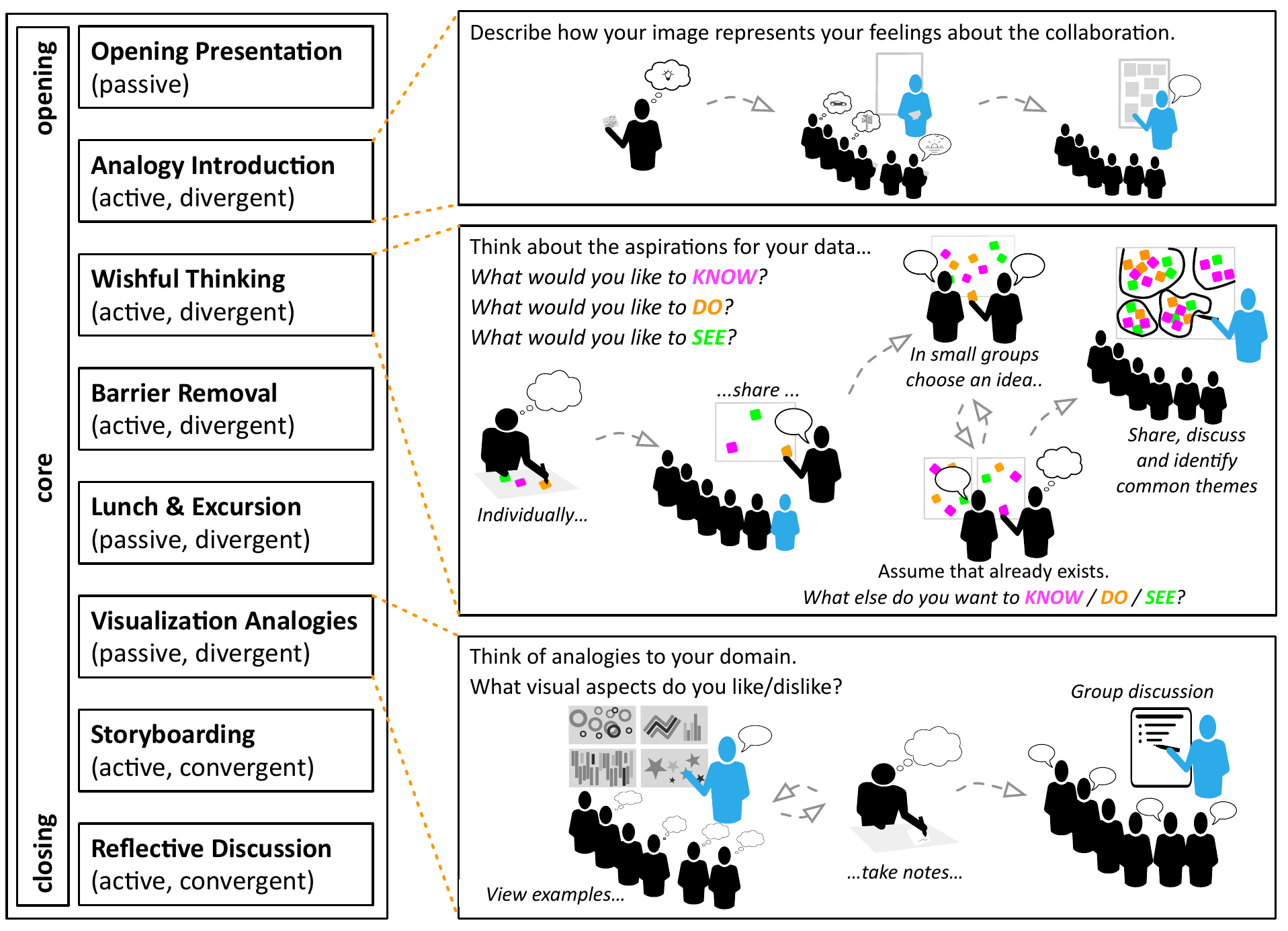}
    \caption{The eight methods of the full-day, example \workshop (left) with the process of three methods summarized graphically (right). The workshop methods diverge to explore a broad ideaspace before they converge to the more promising ideas. Three of the methods are described in the text and the remainder are explained in the Supplemental Material. The methods can be summarized as: 1) the opening presentation establishes creativity principles; 2) an analogy introduction promotes interpersonal leveling; 3) wishful thinking elicits opportunities for visualization; 4) barrier removal explores those opportunities further; 5) lunch \& excursion provides time for rest and incubation; 6) visualization analogies allows specification of requirements by example; 7) storyboarding summarizes key ideas in a graphic form; and 8) the reflective discussion highlights potentially interesting ideas for workshop analysis. This workshop plan is a starting point for future workshops.}
    \label{fig:example-workshop}
\end{figure}

\subsection{Example Workshop \& Methods}
\label{sec:workshop-methods}
To illustrate the workshop structure, we include an example workshop, shown in Fig.~\ref{fig:example-workshop}. We selected this example because it has proven effective in three of our projects [\ref{pro:eon},~\ref{pro:graffinity},~\ref{pro:cp}]. Here, we describe three methods of this workshop that we have also used successfully in additional workshops [\ref{pro:lineage},~\ref{pro:arbor}], and we refer to the Supplemental Material for descriptions of the remaining five methods. We emphasize that this is a starting place for thinking about workshops, and encourage that methods be adopted and adapted for local context.

To explain the workshop methods, we refer to their process --- the steps of execution~\cite{Biskjaer2017}. This process description abstracts and simplifies the methods because during their execution we adapt the process based on participant reactions and our own judgment of the \tactics. 

\subsubsection*{Analogy Introduction} 

We have used this active, interpersonal, and potentially divergent method in the workshop opening. A process of this method, shown in Fig.~\ref{fig:example-workshop} (right, top), starts with a facilitator posing the analogy introduction prompt, e.g., \emph{``If you were to describe yourself as an animal, what would you be and why?''} [\ref{pro:eon}]. The facilitators and participants then respond to the prompt in turn --- expressing themselves creatively. 

Because everyone responds to the eccentric prompt, this method supports interpersonal leveling that helps to develop \trust and \collegiality among stakeholders. Using analogy can prime participants to think creatively~\cite{Gordon1961}.

This method is simple to execute, and participants report that it has a profound impact on the workshop because of the leveling that occurs. The method helps to establish \trust and that all ideas should be accepted and explored [\ref{pro:graffinity}].

A more topical alternative requires more preparation. We have asked participants to come to the workshop with an image that represents their feelings about the project. Participants have created realistic images, clip-art, and sketches to present and discuss [\ref{pro:htva}]. A visual analogy introduction can help establish the \topic of visualization early in the workshop.

\subsubsection*{Wishful Thinking} 

We have used this divergent, active method early in the workshop core. It is based on creativity methods to generate aspirations~\cite{Hicks2004}. We tailored these methods to visualization by prompting participants with a domain scenario and asking questions: {\it ``What would you like to know? What would you like to do? What would you like to see?''}

One process of this method is shown in Fig.~\ref{fig:example-workshop} (right, middle). First, we introduce the prompt and participants answer the know/do/see questions individually on \stickyNotes. Next, participants share ideas in a large group to encourage \collegiality and cross-pollination of ideas. Then, participants form small groups and try to build on their responses by selecting interesting ideas, assuming that they have been completed, and responding to the know/do/see questions again --- increasing the \challenge. Finally, we lead a convergent discussion to highlight interesting ideas and to transition to the next method.

We encourage participants to record answers to the know/do/see questions on different color \stickyNotes because each prompt provides information that is useful at different points in the design process. Participants describe envisaged insights they would like \emph{to know} and analysis tasks that they would like \emph{to do}. Asking what participants would like {\it to see} is often more of a \challenge, but ensures that a \topic of visualization is established early.

We tailor the prompt to the workshop theme and project goals. For example, we asked energy analysts about long term goals for their project --- \emph{``aspirations for the Smart Home programme\ldots''} They generated forward-thinking ideas, e.g., to better understand the value of the data [\ref{pro:eon}]. In contrast, we asked neuroscientists about their current analysis needs --- \emph{``suppose you are analyzing a connectome\ldots''} They created shorter term ideas, e.g., to see neuron connectivity [\ref{pro:graffinity}].


\subsubsection*{Visualization Analogies} 

We have used this divergent, initially passive method later in the workshop core because it promotes incubation while allowing participants to specify visualization requirements by example.  Similar to analogy-based creativity methods~\cite{Gordon1961} and the visualization awareness method~\cite{Koh2011}, we present a curated collection of visualizations and ask participants to individually record analogies to their domain and to specify aspects of the visualizations that they like or dislike. We have used this method repeatedly, iteratively improving its process by reflecting on what worked in a number of our workshops [\ref{pro:edina}~--~\ref{pro:cp},~\ref{pro:arbor}]. 


One process of this method is shown in Fig.~\ref{fig:example-workshop} (right, bottom). First, we provide participants with paper handouts that contain a representative image of each visualization --- we have encouraged participants to annotate the handouts, externalizing their ideas [\ref{pro:graffinity},~\ref{pro:cp},~\ref{pro:arbor}]. Next, we present the curated visualizations on a projector and ask participants to think independently about how each visualization could apply to their domain and record their ideas. Then, we discuss these visualizations and analogies in a large group.

We curate the example visualizations to increase \interest and establish participants' \trust in our visualization expertise. We have used visualizations that we created (to show authority and credibility); those that we did not create (for diversity and to show knowledge of the field); older examples (to show depth of knowledge); challenging examples (to stretch thinking); playful examples (to support engagement and creativity); closely related examples (to make analogies less of a \challenge); and unrelated examples (to promote more challenging divergent thinking).

The discussions during this method have expanded the workshop ideaspace in surprising ways, including \emph{``What does it mean for legends to move?''} [\ref{pro:edina}], \emph{``What does it mean for energy to flow?''} [\ref{pro:eon}], and \emph{``What does it mean for neurons to rhyme?''}~[\ref{pro:graffinity}]. Although this method is primarily passive, participants report that it is engaging and inspiring to see the possibilities of visualization and think about how such visualizations apply to their domain.


\subsubsection*{Additional Methods \& Resources} 

We introduce the example workshop and methods as starting points for future workshops. Yet, the workshop design space is practically infinite and design should be approached with creativity in mind.

To help researchers navigate the design space, our Supplemental Material contains a list of \numberOfExampleMethods example methods that we have used or would consider using in future workshops. For these methods, we describe their process, their influence on the workshop ideaspace, their level of activity, and their potential impact on the \tactics for effective workshops.

We have also found other resources particularly useful while designing workshops. These include books~\cite{CreativeEducationFoundation2015,Gray2010,Hamilton2016,Hohmann2007,Kumar2012,Michalko2006} and research papers~\cite{McFadzean1998,McKenna2014,Sanders2005}. Although these resources target a range of domains outside of visualization, we tailor the workshop methods such that they encourage a visualization mindset and focus on the \topic of visualization opportunities.
\section{During The Workshop: Execute \& Adapt}
\label{sec:during}

Continuing the \workshop process model shown in Fig.~\ref{fig:framework-overview}, we execute the workshop plan. This section proposes five guidelines for workshop execution.

\paragraph{Prepare to execute.} We prepare for the workshop in three ways: resolving details, reviewing how to facilitate effectively, and checking the venue. We encourage researchers to prepare for future workshops in the same ways.

We prepare by resolving many details, such as inviting participants, reserving the venue, ordering snacks for breaks, making arrangements for lunch, etc. Brooks-Harris and Stock-Ward~\cite{Brooks-Harris1999} summarize many practical details that should be considered in preparing for execution. Our additional advice is to promote the visualization mindset in workshop preparation and execution.

We prepare by reviewing principles of effective facilitation, such as acting professionally, demonstrating acceptance, providing encouragement, and using humor~\cite{CreativeEducationFoundation2015,Brooks-Harris1999,Gray2010,Hamilton2016,Stanfield2002}. We also assess our knowledge of the domain because, as facilitators, we will need to lead discussions. Effectively leading discussions can increase \collegiality and \trust between stakeholders as participants can feel that their ideas are valued and understood. In cases where we lacked domain knowledge, we recruited collaborators to serve as facilitators [\ref{pro:cp},~\ref{pro:arbor}]. 

We also prepare by checking the venue for necessary supplies, such as a high quality projector, an Internet connection (if needed), and ample space for group activity. Within the venue, we arrange the furniture to promote a feeling of co-ownership and to encourage \agency~--- a semi-circle seating arrangement works well for this~\cite{Vosko1991}. A mistake in one of our workshops was to have a facilitator using a podium, which implied a hierarchy between facilitators and participants, hindering \collegiality~\cite{Rogers2016}.

\paragraph{Limit distractions.} Workshops provide a time to step away from normal responsibilities and to focus on the \topic. Accordingly, participants and facilitators should be focused on the workshop without distractions, such as leaving for a meeting. 

Communicating with people outside of the workshop --- e.g., through e-mail --- commonly distracts participants and facilitators. It should be discouraged in the workshop opening (e.g., \emph{switch off all electronic devices}). Principles in the workshop opening, however, should be justified to participants. Also, facilitators should lead by example at the risk of eroding \trust and \collegiality.


\paragraph{Guide gently.} While starting execution, the workshop opening can establish an atmosphere in which participants take initiative in completing methods. It is, however, sometimes necessary to redirect the participants in order to stay focused on the \topic. Conversations that deviate from the workshop theme should be redirected. In one workshop [\ref{pro:graffinity}], participants were allowed to discuss ideas more freely, and they reported in feedback that, {\it ``We had a tendency to get distracted [during discussions].''} In a later workshop [\ref{pro:arbor}], we more confidently guided discussions, and participants reported  \emph{\it ``We were guided and kept from going too far off track \ldots this was very effective.''}

However, guiding participants requires judgment to determine whether a conversation is likely to be fruitful. It also requires us to be sensitive to the \tactics~--- e.g., how would redirecting this conversation influence \collegiality or \agency? Redirection can be jolting and can contradict some of the guidelines (e.g., \emph{all ideas are valid}). We can prepare participants for redirection with another guideline during the workshop opening: \emph{Facilitators may keep you on track gently, so please be sensitive to their guidance.} 

\paragraph{Be flexible.} As we guide participants to stay on topic, it is important to be flexible in facilitation. For example, we may spend more time than initially planned on fruitful methods or cut short methods that bore participants. 

Following this guideline can also blur the distinction between participants and facilitators. In one workshop [\ref{pro:htva}], participants proposed a method that was more useful than what was planned. Thus, they became facilitators for this part of the workshop, which reinforced \agency and maintained the \interest of all stakeholders in the project. In the future, we may explore ways to plan this type of interaction, perhaps encouraging participants to create their own methods. 

\paragraph{Adapt tactically.} As we guide the workshop, we interpret group dynamics and adapt methods to the changing situation. We can be forced to adapt for many reasons, such as a failing method (\emph{nobody feels like an animal this morning}; \emph{sticky notes don't stick}), a loss of \interest (\emph{there is no energy}; \emph{the room is too hot}; \emph{we had a tough away day yesterday}); a lack of \agency (\emph{some participants dominate some tasks}); or an equipment failure ({\it projector does not work}; {\it no WiFi connection to present online demos}~[\ref{pro:eon}]). Designing the workshop with alternative methods in mind --- perhaps with varying degrees of \challenge~--- can ensure that workshop time is used effectively.

\paragraph{Record ideas collectively.} Remember: conversations are ephemeral and anything not written down will likely be forgotten. We therefore encourage facilitators and participants to document ideas with context for later analysis. Selecting methods to create physical artifacts can help with recording ideas. As described in Sec.~\ref{sec:design}, externalizing ideas on \stickyNotes and structured prompts has been effective in our workshops and addresses the visualization mindset. 

We are uncertain about the use of audio recording to capture workshop ideas. Although it can be useful for shorter workshops [\ref{pro:lineage}], it can require tremendous time to transcribe before analysis~\cite{Lloyd2011}. Also, recording audio effectively can be challenging as participants move around during the workshop.

It can be useful to ensure that facilitators know that they are expected to help document ideas. A pilot workshop can help with this. In at least one of our projects [\ref{pro:cp}], a pilot workshop may have reduced the note taking pressure on the primary researcher by setting clear expectations that all facilitators should help take notes.


\section{After the Workshop: Analyze \& Act}
\label{sec:after}

After the \workshop, we analyze its output and use the results of that analysis to influence the on-going collaboration. Here, we describe five guidelines for this analysis and action. 

\paragraph{Allocate time for analysis --- soon.} Effective \workshops generate rich and inspiring artifacts that can include hundreds of \stickyNotes, posters, sketches, and other documents. The exact output depends on the methods used in the workshop. Piloting methods can help prepare researchers for the analysis. Regardless, making sense of this output is labor intensive, often requiring more time than the workshop itself. Thus, it is important that we allocate time for analysis, particularly within a day of the workshop, so that we can analyze the workshop output while the experiences are still fresh in our memory.

\paragraph{Create a corpus.} We usually start analysis by creating a digital corpus of the \workshop output. We type or photograph the artifacts, organizing ideas into digital documents or spreadsheets. Through this process, we become familiar with key ideas contained in the artifacts. The corpus also preserves and organizes the artifacts, potentially allowing us to enlist diverse stakeholders --- such as facilitators and collaborators --- in analysis [\ref{pro:graffinity}]. This can help in clarifying ambiguous ideas or adding context to seemingly incomplete ideas.

\paragraph{Analyze with an open mind.} Because the ideas in the workshop output will vary among projects, there are many ways to analyze this corpus of artifacts. We have used qualitative analysis methods --- open coding, mindmapping, and other less formal processes --- to group artifacts into common themes or tasks [\ref{pro:eon},~\ref{pro:graffinity}~--~\ref{pro:updb}]. Quantitative analysis methods should be approached with caution as the frequency of an idea provides little information about its potential importance.

We have ranked the themes and tasks that we discovered in analysis according to various criteria, including novelty, ease of development, potential impact on the domain, and relevance to the project [\ref{pro:eon},~\ref{pro:graffinity}--~\ref{pro:lineage}]. In other cases~[\ref{pro:edina},~\ref{pro:htva}], workshop methods generated specific requirements, tasks, or scenarios that could be edited for clarity and directly integrated into the design process. 

We encourage that analysis be approached with an open mind because of the many ways to make sense of the workshop data, including some approaches that we may not yet have considered.

\paragraph{Embrace results in the visualization design process.} Similarly, \workshop results can be integrated into visualization methodologies and processes in many ways. We have, for example, run additional workshops that explored the possibilities for visualization designs [\ref{pro:edina},~\ref{pro:eon}]. We have applied traditional user-centered design methods, such as interviews and contextual inquiry, to better understand collaborators' tasks that emerged from the workshop~[\ref{pro:graffinity}]. We have created prototypes of varying fidelity, from sketches to functioning software [\ref{pro:graffinity}~--~\ref{pro:lineage}], and we have identified key aims in proposals for funded collaboration~[\ref{pro:arbor}]. 

In all of these cases, our actions were based on the reasons why we ran the workshops, and the workshop results profoundly influenced the direction of our collaboration. For example, in our collaboration with neuroscientists~[\ref{pro:graffinity}], the workshop helped us focus on graph connectivity, a topic that we were able to explore with technology probes and prototypes of increasing fidelity, ultimately resulting in new visualization tools and techniques.

\paragraph{Revisit, reflect, and report on the workshop.} The \workshop output is a trove of information that can be revisited throughout (and even beyond) the project. It can be used to document how ideas evolve throughout applied collaborations. It can also be used to evaluate and validate design decisions by demonstrating that any resulting software fulfills analysis needs that were identified in the workshop data~[\ref{pro:edina}~--~\ref{pro:lineage}]. Revisiting workshop output repeatedly throughout a project can continually inspire new ideas.

In our experience creating this paper, revisiting output from our own workshops allowed us to analyze how and why to use \workshops.  We encourage researchers to reflect and report on their experiences using \workshops, the ways in which workshops influence collaborations, and ideas for future workshops. We hope that this framework provides a starting point for research into these topics.
\section{Discussion}
\label{sec:discussion}

This section discusses implications and limitations of \workshops and the research methodology of critically reflective practice.

\subsection{Limitations of CVO Workshops}

Our experience across diverse domains --- from cartography to neuroscience --- provides evidence that \workshops are a valuable and general method for fostering the visualization mindset while creating artifacts that advance visualization methodologies. We argue that they achieve these goals through the use of \methods that appropriately emphasize the \topic of visualization opportunities while accounting for (inter)personal factors, including \agency, \collegiality, \challenge, \interest, and \trust. 

Yet, workshops may not be appropriate in some scenarios. Because using workshops requires researchers to ask interesting questions and potentially lead discussions about their collaborators' domain, we caution the use of workshops as the first method in a project. Traditional user-centered approaches should be used to learn domain vocabulary and explore the feasibility of collaboration. In the project that did not result in ongoing collaboration~[\ref{pro:updb}], we lacked the domain knowledge needed to effectively design the workshop. Also, our collaborators were too busy to meet with us before the workshop, which should have been a warning about the nature of the project. Accordingly, we recommend researchers evaluate the preconditions of design studies~\cite{Sedlmair2012} in projects where they are considering workshops.

We also recognize that workshops may not be well received by all of the stakeholders. In a full-day workshop [\ref{pro:graffinity}], one participant reported that {\it ``Overall, it was good, but a bit long and slightly repetitive.''} Similarly, after another full-day workshop [\ref{pro:cp}], one participant said {\it ``There was too much time spent expanding and not enough focus \ldots discussions were too shallow and nonspecific.''} Nevertheless, both workshops were generally well received by stakeholders as they allowed us to explore a broad space of visualization opportunities. We can, however, improve future workshops by ensuring that the methods are closely related to the \topic and that we facilitate workshops in a way that provides appropriate \agency to all of the stakeholders.

More generally, whether workshops can enhance creativity is an open question~\cite{Nickerson1999,Sawyer2006}. 
Creativity is a complex phenomenon studied from many perspectives, including design~\cite{Shneiderman2005}, psychology~\cite{Sawyer2006}, sociology~\cite{Lubart1999}, and biology~\cite{Martindale1999}. The results of several controlled experiments indicate that group-based methods can reduce creativity~\cite{Bouchard1969,Mullen1991}. Yet, critics of these studies argue that they rely on contrived metrics and lack ecological validity~\cite{Hewett2005,Mayer1999}. Experimentally testing the relationship between workshops and creativity is beyond the scope of this paper. Instead, we focus on understanding and communicating how we use \workshops in applied collaborations.

\subsection{Critically Reflective Practice}

Throughout this project, we wrestled with a fundamental question: how can we rigorously learn from our diverse, collective experience? We first examined measurable attributes of workshops, such as their length, number of participants, and quantity of ideas generated. However, our workshops were conducted over 10 years in applied settings with no experimental controls. More importantly, it is difficult, if not impossible, to measure how ideas influence collaborations. Quantitative analysis, we decided, would not produce useful knowledge about how to use \workshops. 

We also considered qualitative research methodologies and methods, such as grounded theory~\cite{Corbin1990} and thematic analysis~\cite{Braun2006}. These approaches focus on extracting meaning from externalized data, but the the most meaningful and useful information about workshops resided in our collective, experiential knowledge. We therefore abandoned analysis methods that ignore (or seek to suppress) the role of experience in knowledge generation. 

We found critically reflective practice to be an appropriate approach, providing a methodology to learn from the analysis of experience, documentation, and existing theory, while allowing for the use of additional analysis methods~\cite{Brookfield1998,Thompson2008}. Due to the nature of reflection, however, the framework is not exhaustive, predictive, or objective. Nevertheless, it is consistent with our experience, grounded in existing theory, and, we argue, useful for future visualization research.

Yet, the use of reflective practice may raise questions about the validity of this work. After all, can the framework be validated without experimental data? We emphasize our choice of the term {\it framework}~\cite{Jabareen2008} because we intend for it to be evaluated by whether it provides an interpretive understanding of \workshops. Our position is that it achieves this goal because it enabled us to learn from our experience using workshops on 3 continents over the past 10 years. For example, we used the framework to identify and organize \numberOfPitfalls pitfalls to avoid in future workshops --- they are described in the Supplemental Material. This framework, however, is only a snapshot of our current understanding of \workshops, which will continue to evolve with additional research, practice, and reflection.

Given that this work results from the subjective analysis of our experience, we recognize that there could also be questions about its trustworthiness. Therefore, to increase the trustworthiness of our results, we provide an audit trail~\cite{Carcary2009,Lincoln1985} of our work that contains a timeline of our analysis and our experience as well as diverse artifacts, including comparative analysis of our workshops, presentations outlining the framework, early written drafts of our framework, and structured written reflection to elicit ideas from all of this paper's coauthors. This audit trail, in Supplemental Material, summarizes and includes \numberOfAudits of the reflective artifacts, culled from the original set to protect the privacy of internal discussions and confidential materials from our domain collaborators. 

In future reflective projects, we plan to establish guidelines that encourage transparency of reflective artifacts through  mechanisms to flag documents as on- or off-the-record. Because our research and meta-analysis would have been impossible without well-preserved documentation, we hope that the audit trail inspires future thinking on how to document and preserve the decisions in visualization collaborations. We put forth both the audit trail and our documented use of critically reflective practice as secondary contributions.
\section{Conclusion and Future Work} 
\label{sec:conclusion}

This paper contributes a framework for using workshops in the early stages of applied visualization research. The framework consists of two models for \workshops~--- a process model and a workshop structure. The framework also includes \numberOfGuidelines actionable guidelines for future workshops and a validated example workshop. 

We support the framework with Supplemental Material that includes extended details about the example workshop, \numberOfExampleMethods additional example workshop methods, \numberOfPitfalls pitfalls to avoid in future workshops, and an analysis timeline and audit trail documenting how we developed the framework during a 2-year reflective collaboration. We hope that this framework inspires others to use and report on \workshops in applied visualization research.

Further thinking on the framework reveals opportunities for developing \workshop \methods that emphasize the visualization \mindset. For example, inspired by the Dear Data project~\cite{Lupi2016}, we could ask participants to create graphics that reveal something about their daily life in the week before the workshop. The Dear Data Postcard Kit~\cite{Lupi2017} offers guidance and materials for creating data visualizations about personal experiences, which could be adopted in \workshops.

We also hope to better understand the role of data in \workshops. Visualization methodologies stress the importance of using real data early in collaborative projects~\cite{Lloyd2011,Sedlmair2012}. Our workshops, however, have focused participants on their perceptions of data rather than using real data because working with data is time consuming and unpredictable. In some projects, we incorporated data into the design process by using a series of workshops spaced over weeks or months, providing time for developers to design prototypes between workshops [\ref{pro:edina} -- \ref{pro:htva}]. This development between workshops was expensive in terms of time and effort. But time moves on, and we may be able to reliably use data in workshops with new technologies and techniques, e.g., visualization design tools~\cite{Wongsuphasawat2016}, declarative visualization languages~\cite{Satyanarayan2017}, constructive visualization~\cite{Huron2014}, and sketching~\cite{Walny2015}.

Additionally, in this paper we focused on workshops to elicit visualization opportunities in the early stages of applied work. Exploring how the framework could be influenced by and extended for workshops that correspond to other stages of applied work --- including the creation and analysis of prototypes, the exploration of data, or in the deployment, training and use of completed systems --- may open up opportunities for using creativity in visualization design and research.
\acknowledgments{We are grateful to the participants, facilitators, and fellow researchers in all of our workshops. We thank the following people for their feedback and contributions to this work: the anonymous reviewers, Graham Dove, Tim Dwyer, Peter Hoghton, Christine Pickett, David Rogers, Francesca Samsel, members of the Vis Design Lab at the University of Utah, and members of the giCentre at City, University of London. This work was supported in part by NSF Grant IIS-1350896. }


\bibliographystyle{abbrv}
\bibliography{test,workshops}

\begin{thebibliography}{10}

\bibitem{CreativeEducationFoundation2015}
{\em Creative {{Problem}}-{{Solving Resource Guide}}}.
\newblock {Creative Education Foundation}, Scituate, MA, USA, 2015.

\bibitem{Anderson2000}
L.~W. Anderson, D.~R. Krathwohl, P.~W. Airasian, K.~A. Cruikshank, R.~E. Mayer,
  P.~R. Pintrich, J.~Rathes, and M.~C. Wittrock.
\newblock {\em A {{Taxonomy}} for {{Learning}}, {{Teaching}}, and
  {{Assessing}}: {{A Revision}} of {{Bloom}}'s {{Taxnomy}} of {{Educational
  Objectives}}, {{Abridged Edition}}}.
\newblock {Pearson}, 2000.

\bibitem{Biskjaer2017}
M.~M. Biskjaer, P.~Dalsgaard, and K.~Halskov.
\newblock Understanding creativity methods in design.
\newblock In {\em Proc. {{Conf}}. {{Designing Interactive Syst}}.}, pages
  839--851. {ACM SIGCHI}, 2017.

\bibitem{Bouchard1969}
T.~J. Bouchard.
\newblock Personality, problem-solving procedure, and performance in small
  groups.
\newblock {\em J. Appl. Psychology}, 53(1):1--29, 1969.

\bibitem{Boud1985}
D.~Boud, R.~Keogh, and D.~Walker.
\newblock {\em Reflection: {{Turning Experience}} into {{Learning}}}.
\newblock {Routledge Taylor and Francis Group}, London, UK, 1985.

\bibitem{Braun2006}
V.~Braun and V.~Clarke.
\newblock Using thematic analysis in psychology.
\newblock {\em Qualitative Res. Psychology}, 3(2):77--101, 2006.

\bibitem{Brookfield1998}
S.~Brookfield.
\newblock Critically reflective practice.
\newblock {\em J. of Continuing Edu. in the Health Professions},
  18(4):197--205, 1998.

\bibitem{Brooks-Harris1999}
J.~E. {Brooks-Harris} and S.~R. {Stock-Ward}.
\newblock {\em Workshops: {{Designing}} and {{Facilitating Experiential
  Learning}}}.
\newblock {SAGE Publications, Inc}, Thousand Oaks, CA, USA, 1999.

\bibitem{Buxton2010}
B.~Buxton.
\newblock {\em Sketching {{User Experiences}}: {{Getting}} the {{Design Right}}
  and the {{Right Desing}}}.
\newblock {Morgan Kaufmann}, San Francisco, CA, USA, 2010.

\bibitem{Carcary2009}
M.~Carcary.
\newblock The research audit trail-enhancing trustworthiness in qualitative
  inquiry.
\newblock {\em The Electron. J. of Bus. Res. Methods}, 7(1), 2009.

\bibitem{Corbin1990}
J.~Corbin and A.~Strauss.
\newblock Grounded theory research: Procedures, canons, and evaluative critera.
\newblock {\em Qualitative Sociology}, 13(1):3--21, 1990.

\bibitem{Crotty1998}
M.~Crotty.
\newblock {\em The {{Foundations}} of {{Social Research}}}.
\newblock {SAGE Publications, Inc}, London, UK, 1998.

\bibitem{DeBono1983}
E.~{de Bono}.
\newblock {\em Lateral {{Thinking}} for {{Management}}}.
\newblock {Pelican Books}, Middlesex, UK, 1983.

\bibitem{Dove2014}
G.~Dove and S.~Jones.
\newblock Using data to stimulate creative thinking in the design of new
  products and services.
\newblock In {\em Proc. {{Conf}}. {{Designing Interactive Syst}}.}, pages
  443--452. {ACM SIGCHI}, 2014.

\bibitem{Dove2016}
G.~Dove, S.~Julie, M.~Mose, and N.~Brodersen.
\newblock Grouping notes through nodes: The functions of post-it notes in
  design team cognition.
\newblock In {\em Des. {{Thinking Res}}. {{Symp}}.}, Copenhagen Business
  School, 2016.

\bibitem{Dykes2010}
J.~Dykes, J.~Wood, and A.~Slingsby.
\newblock Rethinking map legends with visualization.
\newblock {\em IEEE Trans. Vis. Comput. Graphics}, 16(6):890--899, 2010.

\bibitem{Goodwin2013}
S.~Goodwin, J.~Dykes, S.~Jones, I.~Dillingham, G.~Dove, D.~Allison,
  A.~Kachkaev, A.~Slingsby, and J.~Wood.
\newblock Creative user-centered design for energy analysts and modelers.
\newblock {\em IEEE Trans. Vis. Comput. Graphics}, 19(12):2516--2525, 2013.

\bibitem{Goodwin2016}
S.~Goodwin, C.~Mears, T.~Dwyer, M.~{Garcia de la Banda}, G.~Tack, and
  M.~Wallace.
\newblock What do constraint programming users want to see? {{Eexploring}} the
  role of visualisation in profiling of models and search.
\newblock {\em IEEE Trans. Vis. Comput. Graphics}, 23(1):281--290, 2016.

\bibitem{Gordon1961}
J.~Gordon, William.
\newblock {\em Synectics - the {{Developmnent}} of {{Creative Capacity}}}.
\newblock {Harper and Row}, New York, NY, USA, 1961.

\bibitem{Gray2010}
D.~Gray, J.~Macanufo, and S.~Brown.
\newblock {\em Gamestorming: A {{Playbook}} for {{Innovators}},
  {{Rulebreakers}}, and {{Changemakers}}}.
\newblock {O'Reilly Media}, Sebastopol, CA, USA, 2010.

\bibitem{Hamilton2016}
P.~Hamilton.
\newblock {\em The {{Workshop Book}}: {{How}} to {{Design}} and {{Lead
  Succesful Workshops}}}.
\newblock {FT Press}, Upper Saddle River, NJ, USA, 2016.

\bibitem{He2017}
S.~He and E.~Adar.
\newblock {{VizItCards}}: A card-based toolkit for infovis design education.
\newblock {\em IEEE Trans. Vis. Comput. Graphics}, 23(1):561--570, 2017.

\bibitem{Hewett2005}
T.~Hewett, M.~Czerwinski, M.~Terry, J.~Nunamaker, L.~Candy, B.~Kules, and
  E.~Sylvan.
\newblock Creativity support tool evaluation methods and metrics.
\newblock In {\em {{NSF Workshop Report}} on {{Creativity Support Tools}}},
  2005.

\bibitem{Hicks2004}
M.~J. Hicks.
\newblock {\em Problem {{Solving}} and {{Decision Making}}: {{Hard}}, {{Soft}},
  and {{Creative Approaches}}}.
\newblock {Thomson Learning}, London, UK, 2004.

\bibitem{Hohmann2007}
L.~Hohmann.
\newblock {\em Innovation {{Games}}: {{Creating Breakthrough Products Through
  Collaborative Play}}}.
\newblock {Addison-Wesley}, Boston, MA, USA, 2007.

\bibitem{Hollis2013}
B.~Hollis and N.~Maiden.
\newblock Extending agile processes with creativity techniques.
\newblock {\em IEEE Software}, 30(5):78--84, 2013.

\bibitem{Horkoff2015}
J.~Horkoff, N.~Maiden, and J.~Lockerbie.
\newblock Creativity and goal modeling for software requirements engineering.
\newblock In {\em Proc. {{Conf}}. {{Creativity}} and {{Cognition}}}, pages
  165--168. {ACM}, 2015.

\bibitem{Huron2016}
S.~Huron, S.~Carpendale, J.~Boy, and J.~D. Fekete.
\newblock Using {{VisKit}}: A manual for running a constructive visualization
  workshop.
\newblock In {\em Pedagogy of {{Data Vis}}. {{Workshop}} at {{IEEE Vis}}},
  2016.

\bibitem{Huron2014}
S.~Huron, S.~Carpendale, A.~Thudt, A.~Tang, and M.~Mauerer.
\newblock Constructive visualization.
\newblock In {\em Proc. {{Conf}}. {{Designing Interactive Syst}}.}, pages
  433--442. {ACM SIGCHI}, 2014.

\bibitem{Jabareen2008}
Y.~Jabareen.
\newblock Building a conceptual framework: Philosophy, definitions, and
  procedure.
\newblock {\em Intern. J. of Qualitative Methods}, 8(4):49--62, 2008.

\bibitem{Jones2008}
S.~Jones, P.~Lynch, N.~Maiden, and S.~Lindstaedt.
\newblock Use and influence of creative ideas and requirements for a
  work-integrated learning system.
\newblock In {\em Int. {{Requirements Eng}}. {{Conf}}.}, pages 289--294.
  {IEEE}, 2008.

\bibitem{Jones2005}
S.~Jones and N.~Maiden.
\newblock {{RESCUE}}: {{An}} integrated method for specifying requirements for
  complex socio-technical systems.
\newblock In J.~L. Mate and A.~Silva, editors, {\em Requirements
  {{Engineering}} for {{Sociotechnical Systems}}}, pages 245--265. {Information
  Resources Press}, Arlington, VA, USA, 2005.

\bibitem{Jones2007}
S.~Jones, N.~Maiden, and K.~Karlsen.
\newblock Creativity in the specification of large-scale socio-technical
  systems.
\newblock In {\em Conf. {{Creative Inventions}}, {{Innovations}} and {{Everyday
  Des}}. {{HCI}}}, 2007.

\bibitem{Kerzner2017:utdb}
E.~Kerzner, {A. Lex}, and M.~Meyer.
\newblock Utah population database workshop (workshop, {U}niversity of {U}tah).
\newblock unpublished, 2017.

\bibitem{Kerzner2017}
E.~Kerzner, A.~Lex, T.~Urness, C.~L. Sigulinsky, B.~W. Jones, R.~E. Marc, and
  M.~Meyer.
\newblock Graffinity: Visualizing connectivity in large graphs.
\newblock {\em Comput. Graph. Forum}, 34(3):251--260, 2017.

\bibitem{Knapp2016}
J.~Knapp, J.~Zeratsky, and B.~Kowitz.
\newblock {\em Sprint: {{How}} to {{Solve Big Problems}} and {{Test}} New
  {{Ideas}} in {{Just Five Days}}}.
\newblock {Simon \& Schuster}, New York, NY, USA, 2016.

\bibitem{Koh2011}
L.~C. Koh, A.~Slingsby, J.~Dykes, and T.~S. Kam.
\newblock Developing and applying a user-centered model for the design and
  implementation of information visualization tools.
\newblock In {\em Proc. {{Int}}. {{Conf}}. {{Inform}}. {{Vis}}.}, pages 90--95.
  {IEEE}, 2011.

\bibitem{Kumar2012}
V.~Kumar and V.~LaConte.
\newblock {\em 101 {{Design Methods}}: A {{Structured Approach}} to {{Driving
  Innovation}} in {{Your Organization}}}.
\newblock {Wiley}, San Francisco, CA, USA, 2012.

\bibitem{Lam2012}
H.~Lam, E.~Bertini, P.~Isenberg, and C.~Plaisant.
\newblock Empirical studies in information visualization: Seven scenarios.
\newblock {\em IEEE Trans. Vis. Comput. Graphics}, 18(9):1520--1536, 2012.

\bibitem{Laural2003}
B.~Laural, editor.
\newblock {\em Design {{Research}}: {{Methods}} and {{Perspectives}}}.
\newblock {MIT Press}, Cambridge, MA, USA, 2003.

\bibitem{Lincoln1985}
Y.~S. Lincoln and E.~Guba.
\newblock {\em Naturalistic {{Inquiry}}}.
\newblock {SAGE Publications, Inc}, Thousand Oaks, CA, USA, 1985.

\bibitem{Lisle2017}
C.~Lisle, E.~Kerzner, {A. Lex}, and M.~Meyer.
\newblock Arbor summit workshop (workshop, {U}niversity of {U}tah).
\newblock unpublished, 2017.

\bibitem{Lloyd2011}
D.~Lloyd and J.~Dykes.
\newblock Human-centered approaches in geovisualization design: Investigating
  multiple methods through a long-term case study.
\newblock {\em IEEE Trans. Vis. Comput. Graphics}, 17(12):2498--2507, 2011.

\bibitem{Lubart1999}
T.~I. Lubart.
\newblock Creativity across cultures.
\newblock In R.~J. Sternberg, editor, {\em Handbook of {{Creativity}}}, pages
  339--350. {Cambridge University Press}, Cambridge, UK, 1999.

\bibitem{Lupi2016}
G.~Lupi and S.~Posavec.
\newblock {\em Dear {{Data}}: {{The Story}} of a {{Friendship}} in
  {{Fifty}}-{{Two Postcards}}}.
\newblock {Penguin}, London, UK, 2016.

\bibitem{Lupi2017}
G.~Lupi and S.~Posavec.
\newblock {\em Dear {{Data Postcard Kit}}: {{For Two Friends}} to {{Draw}} and
  {{Share}} ({{Postcards}})}.
\newblock {Princeton Architectural Press}, New York City, NY, USA, 2017.

\bibitem{Maiden2010}
N.~Maiden, S.~Jones, K.~Karlsen, R.~Neill, K.~Zachos, and A.~Milne.
\newblock Requirements engineering as creative problem solving: A research
  agenda for idea finding.
\newblock In {\em Int. {{Requirements Eng}}. {{Conf}}.}, pages 57--66. {IEEE},
  2010.

\bibitem{Maiden2004}
N.~Maiden, S.~Manning, S.~Robertson, and J.~Greenwood.
\newblock Integrating creativity workshops into structured requirements
  processes.
\newblock In {\em Proc. {{Conf}}. {{Designing Interactive Syst}}.}, pages
  113--122. {ACM SIGCHI}, 2004.

\bibitem{Maiden2007}
N.~Maiden, C.~Ncube, and S.~Robertson.
\newblock Can requirements be creative? {{Experiences}} with an enhanced air
  space management system.
\newblock In {\em Int. {{Conf}}. {{Software Eng}}.}, pages 632--641. {IEEE},
  2007.

\bibitem{Maiden2005}
N.~Maiden and S.~Robertson.
\newblock Developing use cases and scenarios in the requirements process.
\newblock In {\em Proc. {{Intern}}. {{Conf}}. {{Software Eng}}.}, pages
  561--570. {ACM}, 2005.

\bibitem{Marai2018}
G.~E. Marai.
\newblock Activity-centered domain characterization for problem-driven
  scientific visualization.
\newblock {\em IEEE Trans. Vis. Comput. Graphics}, 24(1):913--922, 2018.

\bibitem{Martindale1999}
C.~Martindale.
\newblock Biological bases of creativity.
\newblock In R.~J. Sternberg, editor, {\em Handbook of {{Creativity}}}, pages
  137--152. {Cambridge University Press}, Cambridge, UK, 1999.

\bibitem{Mayer1999}
R.~Mayer.
\newblock Fifty years of creativity research.
\newblock In R.~J. Sternberg, editor, {\em Handbook of {{Creativity}}}, pages
  449--460. {Cambridge University Press}, Cambridge, UK, 1999.

\bibitem{McCurdy2016a}
N.~McCurdy, J.~Dykes, and M.~Meyer.
\newblock Action design research and visualization design.
\newblock In {\em Proc. {{Workshop}} on {{Beyond Time}} and {{Errors}} on
  {{Novel Evaluation Methods}} for {{Vis}}. ({{BELIV}})}, pages 10--18. {ACM},
  2016.

\bibitem{McFadzean1998}
E.~McFadzean.
\newblock The creativity continuum:towards a classification of creative problem
  solving techniques.
\newblock {\em J. of Creativity and Innovation Manage.}, 7(3):131--139, 1998.

\bibitem{McKenna2014}
S.~McKenna, D.~Mazur, J.~Agutter, and M.~Meyer.
\newblock Design activity framework for visualization design.
\newblock {\em IEEE Trans. Vis. Comput. Graphics}, 20(12):2191--2200, 2014.

\bibitem{McKenna2015}
S.~McKenna, D.~Staheli, and M.~Meyer.
\newblock Unlocking user-centered design methods for building cyber security
  visualizations.
\newblock In {\em {{IEEE Symp}}. {{Vis}}. for {{Cyber Security}} ({{VizSec}})},
  2015.

\bibitem{Michalko2006}
M.~Michalko.
\newblock {\em Thinkertoys: A {{Handbook}} for {{Creative}}-{{Thinking
  Techniques}}}.
\newblock {Ten Speed Press}, Emeryville, CA, USA, 2006.

\bibitem{Miller1989}
W.~C. Miller.
\newblock {\em The {{Creative Edge}}: {{Fostering Innovation Where You Work}}}.
\newblock {Basic Books}, New York City, NY, USA, 1989.

\bibitem{Mullen1991}
B.~Mullen, C.~Salas, and E.~Johnson.
\newblock Productivity loss in brainstorming groups: {{A}} meta-analytical
  integration.
\newblock {\em Basic and Appl. Social Psychology}, 12(1):3--23, 1991.

\bibitem{Muller1993}
M.~Muller and S.~Kuhn.
\newblock Participatory design.
\newblock {\em Commun. ACM}, 36(6):24--28, 1993.

\bibitem{Munzner2009}
T.~Munzner.
\newblock A nested model for visualization design and validation.
\newblock {\em IEEE Trans. Vis. Comput. Graphics}, 15(6):921--928, 2009.

\bibitem{Nickerson1999}
R.~S. Nickerson.
\newblock Enhancing creativity.
\newblock In R.~J. Sternberg, editor, {\em Handbook of {{Creativity}}}, pages
  392--430. {Cambridge University Press}, Cambridge, UK, 1999.

\bibitem{Nobre2017}
C.~Nobre, N.~Gehlenborg, H.~Coon, and A.~Lex.
\newblock Lineage: Visualizing multivariate clinical data in genealogy graphs.
\newblock {\em IEEE Trans. Vis. Comput. Graphics, to be published.}, 2018.

\bibitem{Norman1986}
D.~A. Norman and S.~W. Draper.
\newblock {\em User {{Centered System Design}}; {{New Perspectives}} on
  {{Human}}-{{Computer Interaction}}}.
\newblock {L. Erlbaum Associates Inc}, Hillsdale, NJ, USA, 1986.

\bibitem{Osborn1953}
A.~Osborn.
\newblock {\em Applied {{Immagination}}: {{Principles}} and {{Procedures}} of
  {{Creative Problem Solving}}}.
\newblock {Charle Scribener's Sons}, New York, NY, USA, 1953.

\bibitem{Roberts2016}
J.~C. Roberts, C.~Headleand, and P.~D. Ritsos.
\newblock Sketching designs using the five design-sheet methodology.
\newblock {\em IEEE Trans. Vis. Comput. Graphics}, 22(1):419--428, 2016.

\bibitem{Rogers2016}
D.~H. Rogers, C.~Aragon, D.~Keefe, E.~Kerzner, N.~McCurdy, M.~Meyer, and
  F.~Samsel.
\newblock Discovery {{Jam}}.
\newblock In {\em {{IEEE Vis}} ({{Workshops}})}, 2016.

\bibitem{Rogers2017}
D.~H. Rogers, F.~Samsel, C.~Aragon, D.~F. Keefe, N.~McCurdy, E.~Kerzner, and
  M.~Meyer.
\newblock Discovery {{Jam}}.
\newblock In {\em {{IEEE Vis}} ({{Workshops}})}, 2017.

\bibitem{Sakai2015}
R.~Sakai and J.~Aerts.
\newblock Card sorting techniques for domain characterization in problem-driven
  visualization research.
\newblock In {\em Eurographics {{Conf}}. {{Vis}}. ({{Short Papers}})}.
  {Eurographics}, 2015.

\bibitem{Sanders2005}
E.~B.-N. Sanders.
\newblock Information, insipiration, and co-creation.
\newblock In {\em Conf. {{European Academy}} of {{Des}}.}, 2005.

\bibitem{Sanders2010}
E.~B.-N. Sanders, E.~Brandt, and T.~Binder.
\newblock A framework for organizing the tools and techniques of participatory
  design.
\newblock In {\em Proc. {{Participatory Des}}. {{Conf}}.}, pages 195--198,
  2010.

\bibitem{Sanders2008}
E.~B.-N. Sanders and P.~J. Stappers.
\newblock Co-creation and the new landscapes of design.
\newblock {\em CoDesign: Int. J. of CoCreation in Des. and the Arts},
  4(1):5--18, 2008.

\bibitem{Sanders2013}
L.~Sanders and P.~J. Stappers.
\newblock {\em Convivial {{Toolbox}}: {{Generative Research}} for the {{Front
  End}} of {{Design}}}.
\newblock {BIS Publishers}, Amsterdam, The Netherlands, 2013.

\bibitem{Satyanarayan2017}
A.~Satyanarayan, D.~Moritz, and K.~Wongsuphasawat.
\newblock Vega-{{Lite}}: A grammar of interactive graphics.
\newblock {\em IEEE Trans. Vis. Comput. Graphics}, 23(1):341--350, 2017.

\bibitem{Sawyer2003}
K.~R. Sawyer.
\newblock {\em Group {{Creativity}}: {{Music}}, {{Theater}},
  {{Collaboration}}}.
\newblock {Lawrence Erlbaum Associates}, Mahwah, NJ, USA, 2003.

\bibitem{Sawyer2006}
K.~R. Sawyer.
\newblock {\em Explaining {{Creativity}} - the {{Science}} of {{Human
  Innovation}}}.
\newblock {Oxford University Press}, New York, NY, USA, 2006.

\bibitem{Schon1988}
D.~A. Schon.
\newblock {\em The {{Reflective Practitioner}}}.
\newblock {Basic Books}, New York City, NY, USA, 1988.

\bibitem{Sedlmair2010}
M.~Sedlmair, P.~Isenberg, D.~Baur, and A.~Butz.
\newblock Evaluating information visualization in large companies: Challenges,
  experiences and recommendations.
\newblock In {\em Proc. {{Workshop}} on {{Beyond Time}} and {{Errors}} on
  {{Novel Evaluation Methods}} for {{Vis}}. ({{BELIV}})}, pages 79--86. {ACM},
  2010.

\bibitem{Sedlmair2012}
M.~Sedlmair, M.~Meyer, and T.~Munzner.
\newblock Design study methodology: Reflections from the trenches and the
  stacks.
\newblock {\em IEEE Trans. Vis. Comput. Graphics}, 18(12):2431--2440, 2012.

\bibitem{Shneiderman2005}
B.~Shneiderman, G.~Fischer, M.~Czerwinski, and B.~Myers.
\newblock {\em {{NSF Workshop Report}} on {{Creativity Support Tools}}}.
\newblock {National Science Foundation}, 2005.

\bibitem{Shneiderman2006}
B.~Shneiderman and C.~Plaisant.
\newblock Strategies for evaluating information visualization tools.
\newblock In {\em Proc. {{Workshop}} on {{Beyond Time}} and {{Errors}} on
  {{Novel Evaluation Methods}} for {{Vis}}. ({{BELIV}})}, pages 1--7. {ACM},
  2006.

\bibitem{Stanfield2002}
R.~B. Stanfield.
\newblock {\em The {{Workshop Book}}: {{From Individual Creativity}} to {{Group
  Action}}}.
\newblock {New Society Publishers}, Gabriola Island, BC, Canada, 2002.

\bibitem{Thompson2008}
S.~Thompson and N.~Thompson.
\newblock {\em The {{Critically Reflective Practioner}}}.
\newblock {Palgrave Macmillan}, New York, NY, USA, 2008.

\bibitem{Tory2004}
M.~Tory and T.~Moller.
\newblock Human factors in visualization research.
\newblock {\em IEEE Trans. Vis. Comput. Graphics}, 10(1):72--82, 2004.

\bibitem{Vines2013}
J.~Vines, R.~Clarke, and P.~Wright.
\newblock Configuring participation: On how we involve people in design.
\newblock In {\em {{CHI}} '13 {{Proceedings}} of the {{SIGCHI Conference}} on
  {{Human Factors}} in {{Computing Systems}}}, volume~20, 2013.

\bibitem{Vosko1991}
R.~S. Vosko.
\newblock Where we learn shapes our learning.
\newblock {\em New Directions for Adult and Continuing Edu.},
  50(Summer):23--32, 1991.

\bibitem{Walker2013}
R.~Walker, A.~Slingsby, J.~Dykes, K.~Xu, J.~Wood, P.~H. Nguyen, D.~Stephens,
  B.~L.~W. Wong, and Y.~Zheng.
\newblock An extensible framework for provenance in human terrain visual
  analytics.
\newblock {\em IEEE Trans. Vis. Comput. Graphics}, 19(12):2139--2148, 2013.

\bibitem{Walny2015}
J.~Walny, S.~Huron, and S.~Carpendale.
\newblock An exploratory study of data sketching for visual representation.
\newblock {\em Comput. Graph. Forum}, 34(3):231--240, 2015.

\bibitem{Wongsuphasawat2016}
K.~Wongsuphasawat, D.~Motitz, J.~Mackinlay, B.~Howe, and J.~Heer.
\newblock Voyager: Exploratory analysis via faceted browsing of visualization
  recommendations.
\newblock {\em IEEE Trans. Vis. Comput. Graphics}, 22(1):649--658, 2016.

\end{thebibliography}

\end{document}